\pdfoutput=1
\documentclass[12pt]{article}
\usepackage[T1]{fontenc}
\usepackage[utf8]{inputenc}
\usepackage{lmodern}
\usepackage[margin=1in]{geometry}
\usepackage{setspace}
\usepackage{booktabs}
\usepackage{graphicx}
\usepackage{longtable}
\usepackage{array}
\usepackage{enumitem}
\usepackage{amsmath,amssymb}
\usepackage{natbib}
\usepackage{xurl}
\usepackage[colorlinks=true,linkcolor=black,citecolor=black,urlcolor=black]{hyperref}
\usepackage{microtype}
\setlist[itemize]{leftmargin=1.5em}
\setlist[enumerate]{leftmargin=1.8em}
\bibpunct{[}{]}{;}{a}{}{,}
\setcitestyle{aysep={}}
\title{From Authorial Mathematics to Studio Mathematics:\\
Ecobiontic Forms of Proof after Large Language Models}

\author{
Oliver L{\'o}pez-Corona\\[0.5em]
\small Investigadores por M{\'e}xico (IxM),\\
\small Instituto de Investigaciones en Matem{\'a}ticas Aplicadas y en Sistemas (IIMAS),\\
\small Universidad Nacional Aut{\'o}noma de M{\'e}xico (UNAM)
}

\date{}

\begin{document}
\maketitle

\begin{abstract}
Mathematics has often been organized around an authorial subject: one person, or a small group, composing proofs through language, notation, and judgment. Large language models, proof assistants, formal libraries, and repositories now make another production unit technically credible: a human-machine assemblage. This article calls that unit a studio ecobiont and asks when it is epistemically legitimate. Its governance thesis is that human participation is substantive only when the system preserves traceable provenance, reconstructible human competence, capacity to challenge the result, effective authority to stop or withdraw it, and public responsibility. These conditions distinguish a governed studio from a degenerate studio whose human oversight is ceremonial. A comparison of Polymath, the Liquid Tensor Experiment, Danus, and the Jacobian counterexample episode shows that collaboration, formalization, technical orchestration, and epistemic governance are independent dimensions. The proposed understanding audit and contribution-authority trace are governance designs, not validated measures. No causal superiority over authorial practice is claimed.
\end{abstract}

\noindent\textbf{Keywords:} philosophy of mathematics; mathematical practice; proof; understanding; formalization; large language models; proof assistants; Lean; distributed cognition; social epistemology; mathematical authorship

\section{Introduction: the unit of proof-production}

The question raised by large language models in mathematics is often posed in an unhelpful form: can artificial intelligence do mathematics? That question invites either technological triumphalism or defensive skepticism. A more precise philosophical question is this: what can count as the epistemically responsible unit of mathematical production? In much of modern mathematical culture, the default answer has been authorial. A proof is attributed to one person, or to a small group, and its warrant is assessed through a community of competent readers. This authorial arrangement is not the same as private cognition. Mathematical work depends on teachers, seminars, journals, notational traditions, reviewers, libraries, and institutions. Still, the proof as published object has often been organized around an authorial subject.

The authorial picture has good reasons behind it. Mathematical understanding is not exhausted by a derivation. Hardy's account of mathematical creation emphasizes beauty, economy, and permanence \citep{hardy1940}. Lakatos presents mathematical knowledge as a process of conjecture, proof, refutation, and repair \citep{lakatos1976}. Thurston argues that mathematical progress centrally involves communication and understanding, not merely the accumulation of formally correct results \citep{thurston1994}. Rav maintains that proofs carry epistemic content not reducible to the bare existence of formal derivations \citep{rav1999}. Azzouni's derivation-indicator view, Tanswell's criticism of simple dependence accounts, and recent work on rigor, gaps, testimony, and fallibilism show that informal proof remains embedded in practices of judgment, trust, and correction \citep{azzouni2004,tanswell2015,andersen2020gaps,andersen2020testimony,detoffoli2021,burgessdetoffoli2022,tanswell2024}. Any account of post-LLM mathematics that treats proof as merely machine-checkable output loses this background.

Yet the authorial arrangement is not the only possible arrangement. Proof assistants, formal libraries, automated theorem proving, large language models, version-controlled repositories, benchmarks, and software infrastructures now make another arrangement technically credible. A proof project may involve a conceptual lead, an LLM that proposes sketches or tactics, a proof assistant that checks formal derivations, library maintainers who stabilize dependencies, programmers who build tooling, reviewers who evaluate exposition and significance, and institutions that assign credit and responsibility. This article calls such an arrangement \emph{studio mathematics}. The term is not meant to replace ordinary talk of collaboration, formalization, or mechanized proof. It names a particular shift in the unit of production: from the author alone to a governed human-machine assemblage that produces, checks, explains, preserves, and circulates mathematical artifacts.

The contribution is philosophical rather than predictive. I do not argue that future mathematics will be mostly studio mathematics or that LLMs will replace mathematicians. The central claim is narrower: under suitable conditions, the epistemically relevant unit of proof-production can be a governed assemblage of humans, software systems, formal repositories, and validation practices. This unit is infrastructural, formal, computational, and institutionally governed in a stronger sense than ordinary social dependence. The paper makes three claims that must remain separate. Descriptively, some contemporary projects distribute production and warrant across human and technical components. Normatively, such a project is governed only when it preserves substantive human epistemic agency and public responsibility. Causally, whether governed studios improve learning, originality, reliability, or resilience remains unresolved.

\begin{quote}
\textbf{Governance thesis.} A studio is epistemically legitimate as a human mathematical practice only when its human participants retain reconstructible competence, effective challenge and stop rights, traceable responsibility, and access to the dependencies needed for independent scrutiny.
\end{quote}

That task has become urgent because generative systems can break a familiar institutional proxy. Mathematical papers have often functioned not only as containers of results but also as evidence that their authors acquired the understanding, judgment, and craft required to produce them. If a system can generate a publishable derivation while bypassing that acquisition, the paper ceases to certify what the community informally took it to certify. Weinreich calls this a crisis of AI-generated mathematics and argues that outsourcing proof-production threatens the practice through which mathematical understanding is formed \citep{weinreich2026}. The objection is serious. It identifies a proxy rupture, not merely a nostalgic preference for older tools.

The present paper accepts that diagnosis while rejecting one possible inference from it. If mathematical practice is intrinsically valuable, the relevant question is which configurations preserve, distribute, or destroy the work from which understanding emerges. Literature search, counterexample generation, formal verification, proof planning, and complete proof generation are epistemically different interventions. Venkatesh examines how automation may change collective judgments about mathematical importance, while Alper defends experimentation with AI and formalization as mathematical tools \citep{venkatesh2024,alper2026}. Tao describes mathematical AI as a developing ecology of assistance, verification, and collaboration rather than a single replacement event \citep{tao2026age}. The Leiden Declaration gives these distinctions an institutional form by centering disclosure, independent verification, human responsibility, and research autonomy \citep{leiden2026}. Studio mathematics is defensible only if those commitments become operational.

The phrase \emph{CIMA ecobiont} is used for this purpose. CIMA abbreviates four functions: compute, infer, model, and act. An ecobiont is a coupled epistemic assemblage of agents, tools, symbolic media, infrastructures, norms, and validation practices. The term draws from a prior ecobiont ontology rooted in holobiont and niche-construction thinking \citep{margulisfester1991,gilbertsapptauber2012,lopezcorona2019}, but the present use is not biological. It is an ontology of mathematical practice. It is also continuous with distributed cognition and the extended mind, which deny that cognitive agency must be located entirely inside an individual organism \citep{hutchins1995,clarkchalmers1998}. The argument is not that this vocabulary supersedes those literatures. It is that a CIMA ecobiont makes explicit the relation between production, verification, repository memory, public action, and responsibility in post-LLM mathematics.

For mathematics, CIMA has a more specific meaning than generic information-processing. To compute is to manipulate symbolic or formal objects under rule-governed constraints. To infer is to move from premises, constructions, or proof states to warranted consequences. To model is to select definitions, representations, examples, diagrams, formal encodings, and explanatory structures that make a domain tractable. To act is to stabilize mathematical artifacts in public forms: papers, lectures, library contributions, verified files, reusable lemmas, and standards of acceptance. What distinguishes mathematical CIMA is that these functions are constrained by proof, necessity, intelligibility, and transferability across problems.

The structure is as follows. Section 2 clarifies authorial mathematics and the traditional proof subject. Section 3 reviews distributed proof before LLMs. Section 4 introduces the CIMA ecobiont and a claim-relative governance rule. Section 5 defines studio mathematics and compares four cases. Section 6 separates correctness, understanding, explanation, and taste. Section 7 treats credit, authority, testimony, and responsibility. Section 8 gives a limited role to conditional antifragility. Section 9 addresses institutional ecology and protected practice. Section 10 answers objections. Section 11 concludes.

\section{Authorial mathematics and the traditional proof subject}

Mathematics resembles literary labor in one important respect. It can be performed through a relatively light material medium: language, notation, diagrams, memory, and disciplined imagination. A mathematician may work alone for long periods and produce a public artifact whose main medium is symbolic exposition. This is not true of much contemporary science. Particle physics, genomics, climate modeling, and many branches of neuroscience require instruments, data pipelines, large budgets, technicians, and administrative coordination \citep{desollaprice1963,nrc2015}. Mathematics, by contrast, has preserved a strong authorial form.

The authorial form should not be caricatured as the myth of the isolated genius. Even a single-author proof is sustained by a community. It presupposes prior notation, results, informal standards, evaluative habits, referees, and readers. What is authorial is not the whole epistemic life of mathematics, but the production format of many mathematical artifacts. The published proof is ordinarily credited to a bounded authorial unit, and the community treats that unit as responsible for the result's correctness, exposition, and originality.

The authorial form also helps explain why mathematical judgment is partly aesthetic. Beauty, elegance, economy, and conceptual compression are not ornaments added to proof after correctness has been secured. They guide problem choice, definition choice, notation, explanatory route, and standards of satisfactory understanding. Hardy's aesthetic language is therefore not merely romantic; it names a feature of mathematical agency that is difficult to reduce to checking \citep{hardy1940}. Thurston makes a related point when he treats communication of understanding as a central aim of mathematical progress \citep{thurston1994}. A studio regime must therefore be judged not only by whether it can certify derivations, but also by whether it can preserve, redistribute, or distort mathematical taste.

Philosophy of mathematical practice has already complicated the authorial picture. Lakatos shows that proofs develop through criticism and reconstruction, not only through private deduction \citep{lakatos1976}. Thurston emphasizes that the circulation of understanding is a central mathematical aim \citep{thurston1994}. Rav argues that mathematicians prove theorems because proofs display methods, connections, and explanatory information \citep{rav1999}. Andersen shows that published proofs may contain acceptable gaps that are regulated by refereeing and community expectations \citep{andersen2020gaps}. De Toffoli develops a fallibilist account in which mathematical justification is tied to practices of error correction \citep{detoffoli2021}. Avigad argues that mathematical practice supports reliable inference through social and textual mechanisms that approximate formal standards without reducing to them \citep{avigad2021}.

These accounts are important because they show that the traditional proof subject is already socially mediated. The issue is not whether mathematics is social. It is whether the authorial subject remains the only plausible unit for production and responsibility when the work is distributed across formal systems, generative models, repositories, and infrastructure. The studio thesis begins at this point.

\section{Distributed proof before LLMs}

Studio mathematics is not created ex nihilo by large language models. There are at least three prior trajectories.

First, computer-assisted proof has long challenged simple accounts of surveyability and understanding. The four-color theorem raised the question whether a proof too large or computationally dependent to be surveyed in the ordinary way should count as proof \citep{tymoczko1979}. Gonthier's formal proof of the four-color theorem and the machine-checked proof of the odd order theorem show that formal verification can become part of mathematical practice rather than a philosophical curiosity \citep{gonthier2008,gonthier2013}. Hales's work on the Kepler conjecture and the Flyspeck project provide another paradigmatic case in which proof, computation, formalization, and institutional trust became interdependent \citep{hales2008,hales2017}. MacKenzie's sociology of mechanized proof makes the same point at the level of trust and risk: a machine-checked proof is not a disembodied certificate but part of a culture of tools, experts, and conventions \citep{mackenzie2001}.

Second, mathematical collaboration has sometimes exceeded the authorial small-group format. The Polymath projects showed that distributed online collaboration could produce serious mathematics \citep{gowersnielsen2009,polymath2012}. Habgood-Coote and Tanswell analyze the classification of finite simple groups as a case of group knowledge and mathematical collaboration \citep{habgoodcoote2023}. Such cases indicate that a mathematical result can be known, maintained, and warranted by a group with a division of epistemic labor. This is not yet studio mathematics in the post-LLM sense, but it weakens the idea that the only relevant proof subject must be a single authorial agent.

Third, formal libraries and proof assistants have introduced repository-based mathematical memory. Lean 4 supports interactive theorem proving, and mathlib provides a large communal library for Lean \citep{demouraullrich2021,mathlib2020}. The Liquid Tensor Experiment is especially instructive. Its philosophical relevance is not simply that a sophisticated mathematical result was formalized. It is that the work required translation between informal expert understanding and formal dependencies, coordination between mathematicians and formalizers, reliance on a public library ecology, and attention to what the existing library could express \citep{liquidtensor2022,commelin2024}. In such a case, the proof-producing unit is not exhausted by the author of the original argument, the formalizer, the proof assistant, or mathlib considered separately. The unit is a coupled practice in which conceptual mathematics, proof engineering, repository maintenance, and machine checking interact. This is a near case of studio mathematics without needing LLMs as its defining feature.

Large language models enter this already transformed scene. AlphaGeometry, FunSearch, and AlphaProof are not merely isolated technical achievements; they indicate that generation, search, formal language, reinforcement learning, and verification can be coupled in mathematically significant ways \citep{trinh2024,romeraparedes2024,hubert2026}. Benchmarks such as miniF2F and PutnamBench further show that model-assisted mathematical reasoning is increasingly evaluated through shared test environments and formal targets \citep{zheng2022,tsoukalas2024}. These developments matter philosophically because they alter the ecology in which mathematical candidates are generated, filtered, checked, and credited.

\section{CIMA ecobionts as a descriptive ontology}

The CIMA ecobiont is an organizing ontology, not a theorem. It is meant to describe a unit of mathematical production when no single agent, text, or tool is sufficient to explain the result's production and warrant. Let
\[
E = \langle H,T,S,I,R,V,A,\Omega \rangle,
\]
where $H$ is a set of human agents; $T$ a set of technical systems; $S$ the symbolic media; $I$ the institutional environment; $R$ the material and computational resources; $V$ the validation regime; $A$ the public actions and outputs; and $\Omega$ the task environment. An ecobiont exists when these components are coupled around CIMA functions: computation, inference, modeling, and action.

This tuple is not intended to formalize mathematical practice in the strong sense of yielding theorems. It is a disciplined schema for distinguishing production units. A seminar group that discusses a conjecture, an LLM that proposes a proof sketch, and a proof assistant that checks a Lean file are not by themselves the same kind of unit. They differ in coupling, validation, memory, and responsibility. The tuple makes those differences visible.

The schema becomes specifically mathematical only when the validation regime $V$ is tied to proof-sensitive norms. In other domains, validation may be empirical fit, audience response, clinical outcome, or market uptake. In mathematics, $V$ includes derivability, acceptable informal gaps, formal checkability where available, explanatory adequacy, and uptake by competent practitioners. Similarly, $S$ is not any symbolic medium. It includes signs whose use is governed by mathematical constraints: variables, definitions, diagrams, proof states, formal terms, and lemmas with dependency structures. This is why the same CIMA vocabulary can describe different ecobionts without erasing the distinctive normativity of mathematical proof.

\begin{longtable}{p{0.18\linewidth}p{0.35\linewidth}p{0.35\linewidth}}
\caption{CIMA ecobiont components in mathematical production.}\label{tab:cima}\\
\toprule
Component & Role & Example in post-LLM mathematics \\
\midrule
\endfirsthead
\toprule
Component & Role & Example in post-LLM mathematics \\
\midrule
\endhead
$H$ Human agents & Judgment, taste, expertise, responsibility, interpretation. & Mathematicians, formalizers, reviewers, software engineers, students. \\
$T$ Technical systems & Generation, checking, search, storage, automation. & LLMs, Lean, Coq, Isabelle, automated provers, version control. \\
$S$ Symbolic media & Carriers of mathematical form. & Natural language, notation, proof states, formal code, diagrams. \\
$I$ Institutions & Norms, incentives, certification, resources. & Universities, journals, conferences, grants, repositories. \\
$R$ Resources & Conditions of operation. & Compute, libraries, documentation, maintenance labor, time. \\
$V$ Validation & Warranted acceptance. & Peer review, expert reading, formal checking, benchmark evaluation. \\
$A$ Action & Stabilized outputs and uptake. & Papers, library contributions, formal proofs, teaching material. \\
$\Omega$ Task environment & Problem space and constraints. & Conjectures, formalization gaps, library dependencies, open problems. \\
\bottomrule
\end{longtable}

The key distinction is between a tool and a production unit. A proof assistant is a tool. A large language model is a tool. A repository is a tool and an archive. A studio ecobiont is the organized coupling through which these elements jointly produce mathematical outputs. The same distinction helps avoid a common confusion in AI discourse. The question is not whether an LLM understands a proof. The question is what role model output plays in a wider system that includes formal checking, human interpretation, library context, and public accountability.

Human presence alone does not make such a system human-governed. Let $H^{*}\subseteq H$ denote the humans who are authorized and competent to stop release, revise the claim, inspect decisive dependencies, and assume public responsibility. A studio is governed only if $H^{*}$ is non-empty in practice, not merely on an organization chart. A ceremonial human who signs a result but cannot reconstruct its decisive transitions does not supply epistemic governance. This condition becomes the bridge between the descriptive CIMA tuple and the normative account developed below.

Governance is assessed relative to a public claim $q$, rather than assigned once to an entire laboratory or platform. A studio is governed with respect to $q$ only when five conditions hold:

\begin{enumerate}
\item \emph{provenance}: the route from source material, model output, code, formal dependencies, and human decisions to $q$ is inspectable;
\item \emph{reconstruction}: at least one identified member of $H^{*}$ can reconstruct the decisive reasoning at an appropriate level of granularity;
\item \emph{challenge}: $H^{*}$ can diagnose a material error, assess its propagation, and test transfer to a nearby case;
\item \emph{authority}: $H^{*}$ can stop release and can require revision or withdrawal; and
\item \emph{responsibility}: human participants publicly endorse only the claims within their competence, disclose material dependencies, and enable independent scrutiny where licensing and privacy permit.
\end{enumerate}

The conditions are conjunctive for the claim being endorsed. Competence may be distributed across a team, but each decisive dependency must map to an accountable human and the team must be able to integrate those local judgments into a release decision. In this paper, \emph{degenerate} is a technical boundary term for a system that fails this rule while continuing to present its outputs as governed human mathematics.

This ontology also clarifies the relation to existing distributed cognition accounts. Clark and Chalmers argue that parts of the environment may be constitutive of cognitive processes under appropriate conditions \citep{clarkchalmers1998}; Hutchins shows that cognition can be distributed across persons, artifacts, and environments \citep{hutchins1995}. The CIMA ecobiont is narrower. It concerns mathematically relevant production and warrant. Its distinctive task is not to locate cognition in general, but to analyze when proof-production becomes a coupled process of generation, formalization, validation, preservation, and responsibility.

\section{Studio mathematics}

Studio mathematics is a regime of mathematical production in which human expertise, generative systems, formal verification, repository-based memory, software infrastructure, and institutional governance are coupled strongly enough that the product cannot be adequately attributed to an isolated authorial act. The claim is not merely sociological. It concerns the epistemic organization of warrant: how candidate arguments are generated, how errors are detected, how derivations are checked, how explanation is recovered, and how responsibility is assigned. The term \emph{studio} is analogical, but the thesis is literal. The analogy with cinema is useful because cinema distinguishes individual artistic vision from organized production: scripts, actors, cameras, editing, funding, legal clearance, distribution, and reception \citep{becker1982,caves2000,caldwell2008}. Hesse and Bartha emphasize that analogies are valuable when the projected relation is made explicit rather than left as resemblance \citep{hesse1966,bartha2010}. Here the projected relation is division of production labor under a governed regime of validation and public circulation. Mathematics and cinema do not share a truth condition. They may share a production ontology.

\begin{longtable}{p{0.24\linewidth}p{0.30\linewidth}p{0.34\linewidth}}
\caption{Five regimes of mathematical production.}\label{tab:regimes}\\
\toprule
Regime & Typical unit & Main warranting form \\
\midrule
\endfirsthead
\toprule
Regime & Typical unit & Main warranting form \\
\midrule
\endhead
Authorial mathematics & Individual or small group. & Expert reading, exposition, proof tradition, peer review. \\
Collaborative mathematics & Coordinated group or community. & Distributed checking, testimony, shared expertise, public correction. \\
Formal-computational mathematics & Humans plus proof assistants and formal libraries. & Machine checking, formal derivation, library dependency management. \\
Studio mathematics & Governed human-machine ecobiont. & Coupled generation, formal verification, repository memory, expert interpretation, institutional accountability. \\
Degenerate studio mathematics & Output-producing assemblage with ceremonial human oversight. & Performance, authority, or certification without reconstructible human understanding and accountable release. \\
\bottomrule
\end{longtable}

This table marks a difference between studio mathematics and ordinary collaboration. Collaboration can occur without formal systems or reusable repositories. Formalization can occur without LLMs. LLM use can occur without verification. Studio mathematics requires the coupling. It is not mathematics with a chatbot. It is not proof automation alone. It is a production regime in which candidate generation, formal checking, repository memory, human explanation, and public responsibility are integrated. A degenerate studio has much of the same technical surface but lacks the last two properties in a substantive form.

For governance purposes, AI use should also be reported by intervention level rather than by a binary disclosure. The prospective scale runs from $L0$, no generative system used, to $L5$, end-to-end result generation with limited human reconstruction. Intermediate levels distinguish clerical or literature support, local calculation and examples, proof search and decomposition, and substantial proof generation followed by audit. The full scale appears in the supplement. It reports how directly generation enters the justificatory path, not the moral worth of a project.

\subsection{A discriminating comparison of cases}

The framework is useful only if it separates cases that differ in production and warrant. Table~\ref{tab:cases} applies the same criteria to four documented episodes. The classifications concern the publicly described workflows, not the private capacities or intentions of their participants.

\begin{longtable}{@{}p{0.19\linewidth}p{0.31\linewidth}p{0.22\linewidth}p{0.20\linewidth}@{}}
\caption{Comparative classification of mathematical production cases.}\label{tab:cases}\\
\toprule
Case & Coupling and warrant & Evidence about $H^{*}$ & Classification \\
\midrule
\endfirsthead
\toprule
Case & Coupling and warrant & Evidence about $H^{*}$ & Classification \\
\midrule
\endhead
Polymath & Distributed online proof development, public discussion, correction, and conventional mathematical argument \citep{gowersnielsen2009,polymath2012}. & Human contributors supplied the arguments and public checking; no generative system occupied the justificatory path. & Collaborative mathematics, not a studio. \\
Liquid Tensor Experiment & Expert mathematics, Lean formalization, public repository memory, proof engineering, and machine checking \citep{liquidtensor2022,commelin2024}. & Identified mathematicians and formalizers retained authority over translation, dependencies, and release. & Formal-computational near case; it lacks the generative coupling required by the strict studio definition. \\
Danus matroid experiment & Multi-agent generation, fact-graph memory, a verifier, and a paper reported as autonomously produced before the corresponding human paper was public \citep{liu2026danus,cheng2026tangentai,cheng2026tangenthuman}. & The published descriptions establish technical orchestration but do not by themselves establish delayed reconstruction, challenge performance, or collective authorization. & Studio candidate; governance remains undetermined. \\
Jacobian counterexample episode & Model-generated counterexample, computational checking, and an AI-assisted explanatory response, as described by Weinreich \citep{weinreich2026}. & Verification may be possible while the origin and digestion of the idea remain heavily automated; the public record cited here is insufficient to certify the five conditions. & Boundary case illustrating proxy rupture; governance unestablished. \\
\bottomrule
\end{longtable}

The comparison blocks two easy inferences. Distributed authorship does not entail studio mathematics, as Polymath shows. Technical orchestration does not entail governance, as the Danus and Jacobian episodes show. The Liquid Tensor Experiment demonstrates that repository memory, formal verification, and divided labor can deepen warrant without requiring generative AI. Current evidence therefore supports a multidimensional classification, not a historical claim that one regime has already replaced the others. AlphaGeometry, FunSearch, and AlphaProof provide additional evidence that generation, search, formal language, and verification can be coupled in constrained domains \citep{trinh2024,romeraparedes2024,hubert2026}; they do not settle the governance question.

\section{Correctness, understanding, explanation, and taste}

The shift to studio mathematics would be philosophically shallow if it meant only that proofs can be checked by machines. Formal correctness matters, but it does not settle the epistemology of proof. Weatherall and Wolfson argue that formal correctness is neither necessary nor sufficient for a proof's epistemic value \citep{weatherallwolfson2026}. That claim fits a long line of work. Thurston emphasizes understanding. Rav emphasizes the epistemic content of proofs. Azzouni emphasizes the relation between informal proofs and formal derivations. Tanswell argues that dependence relations between informal and formal proofs are not simple. Andersen analyzes acceptable gaps and testimony in ordinary mathematical practice \citep{thurston1994,rav1999,azzouni2004,tanswell2015,andersen2020gaps,andersen2020testimony}.

Studio mathematics therefore needs a layered account of proof. A proof may have at least five layers.

\begin{enumerate}
\item A \emph{generative layer}, where conjectures, examples, proof routes, or formal sketches are proposed.
\item A \emph{formal layer}, where derivations are checked inside a proof assistant or related system.
\item An \emph{expository layer}, where humans explain why the result matters and how the argument works.
\item A \emph{repository layer}, where dependencies, versions, and reusable components are preserved.
\item A \emph{communal layer}, where peers evaluate significance, intelligibility, originality, and trustworthiness.
\end{enumerate}

These layers can come apart. A Lean file may be correct and yet opaque to many mathematicians. An informal proof may be illuminating and yet contain gaps that need filling. An LLM-generated sketch may be suggestive and wrong. A benchmark success may show competence in one domain without supporting broader claims. Hallucination research makes this point vivid: language models can produce fluent outputs that are not warranted by evidence or derivation \citep{huang2025}. The studio response is not to trust models less or more in the abstract. It is to assign each component an epistemic role and to prevent role confusion.

This distinction matters for explanation. If a formalized proof only certifies that a derivation exists, it may fail to explain why a theorem is true or why a method is fruitful. If an informal exposition explains but cannot be checked, it may fail under pressure. A strong studio proof should not collapse explanation into verification. It should preserve both: formal reliability where feasible, and human intelligibility where mathematics requires transmission, teaching, extension, and conceptual use.

It also matters for taste. Mathematical communities do not value all correct proofs equally. They distinguish proofs that merely establish a result from proofs that reveal structure, unify cases, introduce fruitful definitions, or open new lines of inquiry. Studio mathematics can support this judgment when it frees human attention from routine checking or exposes dependency structures that were previously hidden. It can also weaken it if optimization for benchmark success, tactic search, or formal completion displaces the search for conceptual economy. The relevant question is therefore not whether a machine-checked proof is beautiful. It is whether the ecobiont preserves a role for judgments of elegance, depth, and explanatory force.

\subsection{Degenerate studios and the epistemic agency floor}

A \emph{degenerate studio} is a proof-production ecobiont that can generate, certify, or circulate mathematical claims while no accountable human or human group can exercise the relevant epistemic agency over them. Degeneracy is therefore not equivalent to extensive automation. A highly automated project may remain governed if authorized participants can inspect dependencies, reconstruct the decisive reasoning, diagnose failure, and refuse release. Conversely, a nominally human-led project may be degenerate when the human role is restricted to forwarding prompts, accepting certificates, or lending institutional status to outputs that no responsible participant understands.

This definition implies an epistemic agency floor. The proposed \emph{prospective understanding audit} tests four clusters: reconstruction of strategy and decisive transitions; identification of assumptions, imported results, formal dependencies, and generated material; adversarial error diagnosis and transfer to a nearby case; and authority to explain significance, revise the claim, stop release, or withdraw it. The seven-dimension protocol, scoring fields, and administration rules appear in the supplement. The instrument has not been validated as a measure of mathematical understanding, has no validated pass threshold, and will require field-specific calibration. Its present function is governance: it forces a project to name who understands what, which dependencies are delegated, and where responsibility can interrupt the pipeline.

The audit also addresses the indistinguishability problem. In a conversational trace, helping someone understand and saving them the work from which understanding would have emerged may look nearly identical. The distinction becomes more visible only after the conversation, when the participant must reconstruct, diagnose, transfer, and authorize without treating the model transcript as an oracle. The protocol cannot guarantee understanding, but it makes ceremonial oversight harder to confuse with epistemic agency.

The proposal is intentionally prospective. A future benchmark could compare authorial, assistive, governed-studio, and degenerate-studio workflows on retention, error diagnosis, transfer, originality, and time. Until such a benchmark exists, this paper makes no causal claim that governed studios outperform authorial practice. The present contribution is to define the comparison and its failure conditions.

\section{Credit, testimony, and responsibility}

A change in the unit of production creates a change in the unit of credit. Traditional authorship already struggles with large collaborations, software, data, and infrastructure. In studio mathematics the problem becomes more acute. A result may involve conceptual design, problem decomposition, model prompting, proof search, formalization, library maintenance, software engineering, review, documentation, and exposition. Treating all of this as either ``authorship'' or ``acknowledgment'' is too crude.

The social epistemology of mathematics is directly relevant. Andersen, Andersen, and S{\o}rensen argue that testimony plays a role in mathematics \citep{andersen2020testimony}. Habgood-Coote and Tanswell analyze group knowledge in the classification of finite simple groups \citep{habgoodcoote2023}. Rittberg, Tanswell, and Van Bendegem show that epistemic injustice can occur in mathematical practice \citep{rittberg2020}. Work on authorship in philosophy of science further shows that author lists are not merely decorative: they distribute responsibility, recognition, and authority \citep{habgoodcoote2024authorship}.

Studio mathematics should therefore distinguish at least five kinds of contribution.

\begin{enumerate}
\item \emph{Conceptual contribution}: posing the problem, choosing definitions, identifying the strategy.
\item \emph{Generative contribution}: producing candidate lemmas, sketches, examples, or search paths.
\item \emph{Formal contribution}: encoding and checking derivations in a proof assistant.
\item \emph{Infrastructural contribution}: maintaining libraries, tools, datasets, environments, and documentation.
\item \emph{Interpretive contribution}: explaining, simplifying, evaluating, and contextualizing the result.
\end{enumerate}

These categories need not map mechanically onto authorship positions. They provide a vocabulary for responsibility. If model-generated content materially shaped a proof route, that use should be documented. If library maintainers made a result possible through substantial formal infrastructure, that labor should be visible. If a proof depends on a proprietary system that others cannot inspect, the epistemic status of the claim differs from that of an openly checkable artifact. The aim is not bureaucracy. The aim is to align credit with epistemic labor.

Contribution and epistemic authority should therefore be represented on separate axes. For participant $i$, let
\[
C_i=(P_i,E_i),
\]
where $P_i$ records productive contributions and $E_i$ records the scope of claims that the participant is competent and authorized to endorse. A library maintainer may have high productive importance without claiming authority over the theorem's significance. A conceptual lead may have broad interpretive authority while relying on formalizers for local certificates. A model may be causally productive without being a bearer of testimony or responsibility. The two-axis representation prevents the false choice between granting conventional authorship to every enabling contribution and making infrastructure invisible.

Weinreich's suggestion that ideas might be co-owned by anyone who demonstrates authoritative understanding points toward a valuable correction to paper-centered authorship \citep{weinreich2026}. It treats understanding as something to be demonstrated rather than inferred from a byline. Yet understanding cannot be the only source of credit. Formalization, maintenance, dataset construction, and software engineering may deserve recognition even when their contributors do not claim authoritative understanding of the whole result. The dual-axis model retains the insight while making room for infrastructural labor.

A minimal governance standard would require a contribution and authority trace, not merely a final author list. Such a trace would record which parts of a proof were proposed informally, generated computationally, checked formally, imported from existing libraries, modified during review, or dependent on non-public systems. It would also identify who may authorize each major claim and who can stop release. The trace need not be exhaustive. It should be sufficient for competent readers to understand the epistemic route from conjecture to public artifact. This also helps distinguish credit from responsibility: a contributor may deserve credit for infrastructure without being responsible for the theorem's exposition, and a conceptual author may remain responsible for claims of significance even when formal checking is delegated.

\section{Conditional antifragility}

The vocabulary of antifragility should be used carefully. Studio mathematics is not automatically better because it is distributed, formal, or model-assisted. It may be fragile. It may concentrate power in groups with access to compute, proprietary models, and engineering support. It may create brittle dependencies. It may encourage trust inflation, where formal-looking artifacts receive more authority than they deserve. It may also create maintenance burdens that are poorly rewarded.

A precise use of antifragility avoids this problem. Let
\[
\{S,\mathcal{P},U\}
\]
name a system, a class of perturbations, and a utility measure. In this setting, $S$ is a proof-production ecobiont; $\mathcal{P}$ includes counterexamples, failed formalizations, hallucinated steps, broken dependencies, adversarial peer review, benchmark failures, automation shocks, rapid mass generation, deskilling pressure, and proprietary lock-in; $U$ includes reliability, intelligibility, retained human understanding, epistemic agency, institutional independence, documentation quality, library robustness, reusable components, and fair attribution. Following resilience and antifragility work \citep{holling1973,taleb2012,monperrus2017,equihua2020,axenie2023}, studio mathematics is antifragile only when perturbations in $\mathcal{P}$ systematically improve $U$.

This condition is demanding. A failed formalization is useful only if it reveals a gap, improves a definition, strengthens a library, or documents a limitation. An LLM hallucination is useful only if the system can detect it and use it as negative evidence or a prompt for refinement. Peer review is useful only if it improves correctness and understanding rather than merely imposing conservatism. A benchmark failure is useful only if it updates models, methods, or claims. Conditional antifragility is therefore a normative design principle: build proof ecobionts that learn from disorder without mistaking disorder for progress.

Four operational conditions follow. First, perturbations must be retained: failures, gaps, and dependency breaks should leave inspectable traces. Second, perturbations must be selected: not every failure is informative, and the system needs criteria for distinguishing noise from useful stress. Third, perturbations must be converted: a detected gap should become a corrected proof, a better lemma, a library improvement, a benchmark update, or a revised claim. Fourth, conversion must be public enough to support trust: if only a private model or closed platform learns from the failure, the mathematical community may not gain the relevant utility. On this reading, antifragility is not a property of LLM use. It is a property of governed error conversion.

Ashby's law of requisite variety helps explain why this matters \citep{ashby1956}. A system can regulate disturbances only if it has enough internal variety. In studio mathematics, variety comes from combining human experts, formal systems, repositories, independent checkers, open review, and maintenance practices. But variety without governance can produce noise. The design problem is to create enough heterogeneity to detect error, enough structure to preserve trust, and enough openness to correct the system.

\section{Institutional ecology and protected mathematical practice}

The crisis objection is institutional as well as epistemic. Public support for mathematicians is easier to defend when mathematical workers are visibly necessary to discovery, verification, teaching, and stewardship. If institutions treat machine-generated output as a substitute for mathematical employment, then even a technically successful system can damage the community that makes its outputs intelligible and trustworthy. Weinreich is right that this cannot be answered by saying that mathematicians may simply choose to keep practicing for intrinsic reasons \citep{weinreich2026}. Time, positions, journals, and training environments are material conditions of that practice.

Total opposition, however, does not follow. If high-capability systems diffuse regardless of individual abstention, a single prohibition can turn refusal into an exclusion mechanism while leaving the surrounding political economy unchanged. The more defensible response is a plural institutional ecology with protected authorial practice, governed studio practice, and explicit barriers against degenerate production. The Leiden Declaration supplies an emerging community-level commitment to responsible AI use, while Tao's account of mathematical AI emphasizes adaptation of norms and workflows rather than a single technological destiny \citep{leiden2026,tao2026age}.

Protected practice can take several forms. Journals and conferences can create disclosure classes rather than a binary AI label. Hiring and training can value oral reconstruction, error diagnosis, conceptual contribution, formalization, and maintenance instead of treating paper counts as a complete proxy. Funders can support open proof infrastructure and compute access so that studio mathematics does not become proprietary by default. Departments can preserve spaces in which students solve problems without generative assistance, because formative practice and production efficiency are different goods. Communities can also maintain venues for explicitly authorial mathematics. These arrangements are not attempts to stop technical development. They are ways of keeping mathematical agency, diversity of practice, and public accountability inside the institutional objective function.

No single governance regime should be universal. Some domains will remain primarily authorial; some formalization projects will be collective but not generative; some discovery programs will become studio-like. The relevant red line is not tool use but degeneracy: acceptance of results whose human sponsors cannot meet the applicable epistemic agency floor. That boundary also clarifies why the economic and existential-risk debates should not be conflated. Catastrophic AI risk may be important, but it requires different premises and policies. The present argument concerns the nearer institutional design of mathematical practice.

Formal Concept Analysis remains available as an optional representational tool in the supplement \citep{ganterwille1999}. It maps which attributes close the proposed taxonomy, but it is neither evidence for the thesis nor a premise of the institutional argument. The supplement also provides a claim ledger, a robustness and heterogeneity map, and figure files. The main philosophical case is text-centered and survives without them.

\section{Objections}

\emph{Objection 1: Studio mathematics is just collaboration under a new name.} Collaboration is part of the story, but not the whole story. Studio mathematics requires a specific coupling of generative systems, formal verification, repository memory, human interpretation, and institutional responsibility. A collaborative seminar without those features is not studio mathematics. A solo proof formalized in Lean may be formal-computational without being studio-like. The studio category is narrower than sociality and broader than automation.

\emph{Objection 2: The cinema analogy is unnecessary.} The analogy is not load-bearing. It is heuristic. It helps mark the difference between authorial production and coordinated production. The literal claim is about proof-production units: a governed human-machine assemblage can become epistemically relevant in ways that an individual author alone is not. The argument would survive without the cinema analogy.

\emph{Objection 3: Formal proof already solved the problem.} Formal proof solves one problem: checking derivations relative to a formal system. It does not by itself solve explanation, significance, surveyability, testimony, authorship, or responsibility. That is why a studio proof must be layered rather than merely certified.

\emph{Objection 4: LLMs are unreliable and should not be part of proof practice.} LLM outputs are unreliable when treated as warranted mathematical assertions. They may still be useful as generators of candidates, examples, search paths, or expository drafts, provided their status is recorded and their outputs are checked. The framework distinguishes generation from verification precisely to prevent unwarranted trust.

\emph{Objection 5: Authorial mathematics remains superior for deep understanding.} It may be superior in many cases. The thesis is not replacement. It is pluralization. Authorial mathematics, collaborative mathematics, formal-computational mathematics, and studio mathematics can coexist. The value of each depends on the mathematical task, the standards of explanation, and the available validation practices.

\emph{Objection 6: Studio mathematics threatens mathematical beauty.} It can. A production regime optimized for formal completion, tactic success, or benchmark performance may marginalize elegance and explanatory economy. But the same regime can also support aesthetic judgment by exposing dependencies, eliminating routine uncertainty, and freeing attention for conceptual organization. The issue is not whether studio mathematics is intrinsically beautiful or ugly. It is whether its governance preserves space for mathematical taste as an epistemic virtue.

\emph{Objection 7: The chess intermediary has not played the game.} Weinreich's analogy is decisive against one weak notion of achievement: a novice who merely relays moves between stronger players has not thereby defeated a grandmaster \citep{weinreich2026}. The same holds for a person who submits model output without understanding it. This is precisely the degenerate-studio case. But the analogy does not cover every distributed practice. A director, formalizer, reviewer, or library maintainer may contribute neither the first move nor the final certificate while still exercising skilled judgment that changes the mathematical product. The contribution and authority trace is meant to separate these roles from mere intermediation.

\emph{Objection 8: If the model can also digest and explain the proof, the human role disappears.} Model-assisted exposition can indeed undermine a proposal that reserves ``digestion'' as the final human monopoly. The studio account therefore does not define human agency by an unautomated task. It defines it by reconstructible competence, authority to intervene, and responsibility for public action. A model may help a participant acquire that competence; the audit asks whether the competence persists when challenged by diagnosis and transfer. If no human can meet the floor, the result may still be formally correct, but it should not be represented as a governed human mathematical achievement.

\emph{Objection 9: Governance will not prevent labor displacement or institutional capture.} Correct. A contribution trace and an understanding audit are not labor policy. They can make displacement and dependence visible, but they cannot by themselves preserve jobs, distribute compute, or constrain platforms. That is why the institutional ecology requires protected training, open infrastructure, plural venues, and evaluation practices that do not reward output volume alone. These proposals are partial safeguards, not a prediction that the transition will be benign.

\section{Conclusion}

The post-LLM transformation of mathematics is not best understood as the arrival of artificial mathematicians. It is better understood as a change in the possible unit of proof-production. Mathematics can still be authored in a literary mode: a person or small group develops a proof through language, notation, judgment, and style. That mode remains central to mathematical understanding. But another mode is now technically credible. In studio mathematics, proof-production is distributed across human expertise, generative systems, proof assistants, formal libraries, repositories, and institutions. The relevant distinction is therefore not authorial versus automated mathematics, but authorial, governed studio, and degenerate studio practice.

The CIMA ecobiont ontology provides a disciplined way to describe this shift. It does not deny existing accounts of mathematical practice, distributed cognition, formal proof, or group knowledge. It builds on them to analyze a new coupling of generation, formal verification, repository memory, explanation, credit, and responsibility. Its mathematical specificity lies in the fact that this coupling remains constrained by proof, necessity, intelligibility, and taste. The epistemic agency floor, the prospective understanding audit, and the contribution-authority trace specify what ``human-governed'' must mean if it is to be more than a slogan.

The appropriate conclusion is plural rather than triumphalist or prohibitionist. Communities should preserve authorial practice, develop governed studios where their benefits are real, and refuse to represent degenerate output pipelines as understood human mathematics. Whether governed studios improve learning, originality, reliability, or resilience remains an empirical question. The paper has defined the comparison, not settled it. The philosophical issue is how mathematical agency, trust, understanding, and judgment should be organized when proof becomes a collective, formal, verifiable, and infrastructural artifact.

\section*{Tool and computational resource disclosure}

Generative AI systems were used during development of this manuscript for literature triage, argument mapping, language revision, LaTeX editing, and generation of reproducibility scripts. The human author defined the research question and framework, selected the claims advanced, and retains responsibility for the interpretations, references, classifications, and final text. No claim is treated as established because it was generated or endorsed by a model. The supplementary package records claim status, source checks, decision rules, and the limits of the unexecuted validation program.

\clearpage
\appendix
\setcounter{section}{0}
\setcounter{figure}{0}
\setcounter{table}{0}
\renewcommand{\thesection}{S\arabic{section}}
\renewcommand{\thefigure}{S\arabic{figure}}
\renewcommand{\thetable}{S\arabic{table}}
\renewcommand{\theHsection}{S.\arabic{section}}
\renewcommand{\theHfigure}{S.\arabic{figure}}
\renewcommand{\theHtable}{S.\arabic{table}}
\singlespacing
\setlength{\parskip}{0.35em}
\setlength{\parindent}{0pt}
\setlength{\tabcolsep}{3pt}

\begin{center}
{\LARGE\bfseries Supplementary Information\par}
\vspace{0.6em}
{\large MIAT-MD audit, governance protocols, and optional representations\par}
\vspace{0.3em}
{\normalsize Version 2.6.1\par}
\end{center}
\bigskip

\section{Status, scope, and nonclaims}

This supplement exposes the inferential architecture of version 2.6.1. It separates four layers: \textbf{ontology} (what entities and regimes are being defined), \textbf{epistemology} (what counts as correctness, understanding, authority, and responsibility), \textbf{methodology} (how those distinctions could be audited), and \textbf{application} (what the framework suggests for current systems and institutions). It also records the comparative case matrix and the staged validation program. A definition is not treated as evidence, a diagram is not treated as a result, and a model output is not promoted to an inference merely because it is fluent or formally shaped.

The understanding audit, AI-use scale, contribution-authority trace, and proposed benchmark are prospective. They have not been validated psychometrically or causally. The FCA and robustness outputs are representations computed from a disclosed hand-coded context. They do not establish that the ontology is true. No claim is made that governed studios outperform authorial practice.

\section{Evidence contract}

Every substantive claim receives a stable ID and records the following fields:

\begin{longtable}{p{0.19\linewidth}p{0.73\linewidth}}
\toprule
Field & Function \\
\midrule
Claim ID & Stable identifier used across manuscript, ledgers, scripts, and audit outputs. \\
Layer & Ontology, epistemology, methodology, or application. \\
Claim type & Definitional, literature-supported, conceptual, normative, empirical example, or prospective hypothesis. \\
Evidence basis & Source, argument, example, or explicit statement that evidence is absent. \\
Validation route & What could check, challenge, or falsify the claim at the appropriate level. \\
RHM status & \texttt{ROBUST\_CORE}, \texttt{SPURIOUS\_ENVELOPE}, or \texttt{LEGITIMATE\_OPEN\_UNCERTAINTY}. \\
Uncertainty & Low, medium, or high, interpreted qualitatively. \\
POST\_HOC & Marker for analyses or criteria introduced after inspecting outcomes. No claim in this release is marked post hoc because no outcome benchmark has been run. \\
\bottomrule
\end{longtable}

\section{AI-use disclosure scale}

The scale reports where generative systems enter the justificatory path. It does not rank ethical worth and should be accompanied by a plain-language tool disclosure.

\begin{longtable}{p{0.08\linewidth}p{0.29\linewidth}p{0.50\linewidth}}
\toprule
Level & Intervention & Minimum disclosure \\
\midrule
$L0$ & No generative system used. & State whether conventional software, search, or proof assistants were used. \\
$L1$ & Clerical or literature support. & Tool, date/version when available, and tasks such as search, formatting, or language editing. \\
$L2$ & Local calculation or example support. & Prompts or procedures material to examples, code, calculations, or counterexamples; independent checks. \\
$L3$ & Proof search or decomposition. & Candidate lemmas, search paths, discarded suggestions, and verification route. \\
$L4$ & Substantial proof or formal derivation generation followed by audit. & Generated sections, formal dependencies, audit record, responsible human, and stop authority. \\
$L5$ & End-to-end result generation with limited human reconstruction. & Full provenance and an explicit warning that the epistemic agency floor has not been met. \\
\bottomrule
\end{longtable}

\section{Operational governance rule}

Governance is classified relative to an endorsed public claim $q$. A project is governed with respect to $q$ only when all five conditions below are documented. A laboratory, platform, or author list does not receive a permanent global classification.

\begin{longtable}{p{0.10\linewidth}p{0.23\linewidth}p{0.58\linewidth}}
\toprule
ID & Condition & Observable record \\
\midrule
GR1 & Provenance & Sources, model outputs, code, formal dependencies, and material human decisions can be traced to the released claim. \\
GR2 & Reconstruction & At least one identified member of $H^{*}$ can reconstruct the decisive reasoning at field-appropriate granularity. \\
GR3 & Challenge & $H^{*}$ can diagnose a material error, trace its effects, and test transfer to a nearby case. \\
GR4 & Authority & $H^{*}$ has an effective right to stop release and require revision or withdrawal. \\
GR5 & Responsibility & Human participants delimit the claims they endorse, disclose material dependencies, and enable independent scrutiny where licensing and privacy permit. \\
\bottomrule
\end{longtable}

Classification is conjunctive for the claim under review. Distributed competence is permitted, but every decisive dependency must map to an accountable human, and the group must demonstrate an integration path from local judgments to the release decision. Failure of a condition classifies the workflow as ungoverned with respect to that claim. The term \emph{degenerate studio} is reserved for an ungoverned workflow that nevertheless represents its output as governed human mathematics.

\section{Claim ledger}

The machine-readable ledger is \texttt{data/claims.csv}. The compact view below records the main claim and its present status.

\begin{longtable}{p{0.07\linewidth}p{0.16\linewidth}p{0.52\linewidth}p{0.17\linewidth}}
\toprule
ID & Layer & Claim & RHM status \\
\midrule
\endfirsthead
\toprule
ID & Layer & Claim & RHM status \\
\midrule
\endhead
C01 & Ontology & CIMA tuple represents a proof-production ecobiont. & Robust core \\
C02 & Ontology & Studio mathematics is a regime, not mere LLM use. & Robust core \\
C03 & Ontology & Degenerate studios have output capability without substantive human governance. & Robust core \\
C04 & Epistemology & Correctness is not sufficient for understanding or epistemic value. & Robust core \\
C05 & Epistemology & Human presence is insufficient without reconstructive authority. & Robust core \\
C06 & Epistemology & Conversation alone may not distinguish learning from bypass. & Robust core \\
C07 & Methodology & An understanding audit can operationalize an agency floor. & Open uncertainty \\
C08 & Methodology & Disclosure should use intervention levels rather than a binary label. & Robust core \\
C09 & Methodology & Contribution and epistemic authority require separate axes. & Robust core \\
C10 & Application & Fact-graph systems demonstrate technical orchestration. & Robust core \\
C11 & Application & Orchestration does not entail understanding or responsibility. & Robust core \\
C12 & Application & Governed studios outperform authorial workflows. & Open uncertainty \\
C13 & Application & Plural institutions better preserve agency than uniform regimes. & Open uncertainty \\
C14 & Application & Protected authorial practice remains valuable. & Robust core \\
C15 & Methodology & FCA is optional representation, not evidence. & Robust core \\
C16 & Methodology & RHM separates robust, spurious, and legitimately open claims. & Robust core \\
C17 & Methodology & Governance requires provenance, reconstruction, challenge, authority, and responsibility for each endorsed claim. & Robust core \\
C18 & Application & A common case matrix separates collaboration, formalization, orchestration, and governance. & Robust core \\
C19 & Methodology & Corpus-scale MIAT-MD requires a frozen codebook and validated gold set before scaling. & Robust core \\
\bottomrule
\end{longtable}

The label ``robust core'' means that a claim survives the distinctions and challenges represented in this audit. It does not mean the claim has been empirically proven. In this release 16 claims are in the robust core, three are legitimate open uncertainties, and none is assigned to the spurious envelope. The empty spurious class is a current audit result, not a guarantee against future demotion.

\section{Comparative case matrix}

The machine-readable table is \texttt{data/case\_matrix.csv}. ``Unreported'' means that the cited public record does not establish the condition; it does not assert that the relevant competence or authority was absent in private.

\begin{longtable}{p{0.20\linewidth}p{0.19\linewidth}p{0.18\linewidth}p{0.17\linewidth}p{0.18\linewidth}}
\toprule
Case & Generative coupling & Formal or repository coupling & Evidence about $H^{*}$ & Classification \\
\midrule
Polymath \cite{gowersnielsen2009,polymath2012} & None in the documented workflow. & Public discussion and conventional proof record. & Strong public evidence of human reasoning and correction. & Collaborative, not studio. \\
Liquid Tensor Experiment \cite{liquidtensor2022} & None required by the documented workflow. & Lean, mathlib, public dependencies, and proof engineering. & Identified experts controlled translation, formalization, and release. & Formal-computational near case. \\
Danus matroid experiment \cite{liu2026danus,cheng2026tangentai,cheng2026tangenthuman} & Multi-agent generation with fact-graph memory. & Stateless verification and stored proof dependencies. & Reconstruction, challenge performance, and collective authorization are unreported. & Studio candidate; governance undetermined. \\
Jacobian episode \cite{weinreich2026} & Generated counterexample and AI-assisted digestion as reported by Weinreich. & Computational checking reported; public provenance remains incomplete in the source used here. & The five governance conditions are not established by the cited record. & Boundary case; governance unestablished. \\
\bottomrule
\end{longtable}

\section{Hypothesis tournament}

Competing possibilities remain live until a discriminating design is executed.

\begin{longtable}{p{0.06\linewidth}p{0.30\linewidth}p{0.31\linewidth}p{0.25\linewidth}}
\toprule
ID & Hypothesis & Discriminating prediction & Failure condition \\
\midrule
H1 & Authorial units remain sufficient for most advanced mathematics. & Tooling changes efficiency but not the responsible unit. & Persistent cases lack any bounded authorial warrant. \\
H2 & Governed studios form a distinct regime. & Stable role differentiation, memory, and collective authorization emerge. & Cases reduce to collaboration or tool use. \\
H3 & Degenerate studios can certify output without retained understanding. & Correctness coexists with failures of reconstruction, diagnosis, and transfer. & Participants retain comparable understanding. \\
H4 & Governance preserves agency under substantial automation. & Authority traces and stop rights predict better diagnosis and responsibility. & Governance labels do not predict behavior. \\
H5 & Plural institutions are more robust than uniform adoption or refusal. & Mixed regimes preserve diversity and correction capacity under shocks. & Mixed regimes increase exclusion or fragility. \\
H6 & MIAT-MD can be applied reliably across arXiv after calibration. & Independent coders and the automated pipeline retain acceptable agreement across fields and time. & Agreement collapses under domain shift or key variables cannot be recovered from papers. \\
\bottomrule
\end{longtable}

\section{MIAT-MD gates G0--G6}

\begin{longtable}{p{0.08\linewidth}p{0.19\linewidth}p{0.48\linewidth}p{0.16\linewidth}}
\toprule
Gate & Focus & Current evidence & Result \\
\midrule
G0 & Scope & Object, comparator, and prospective status are explicit. & Pass \\
G1 & Ontology & CIMA tuple, $H^{*}$, governed studio, degenerate studio, and the claim-relative governance rule are defined. & Pass \\
G2 & Epistemology & Correctness, understanding, explanation, taste, authority, and responsibility remain distinct. & Pass \\
G3 & Methodology & $L0$--$L5$, audit criteria, dual-axis ledger, and nonclaims are specified. & Pass with limitation \\
G4 & Evidence & Claim IDs link source, warrant, validation route, and uncertainty. & Pass \\
G5 & Robustness & H1--H6, RHM classes, comparative cases, and optional FCA variants are disclosed. & Pass with limitation \\
G6 & Governance & Authority, stop rights, release, and withdrawal are specified. & Pass with limitation \\
\bottomrule
\end{longtable}

The limitations at G3, G5, and G6 are substantive: the instruments have not been validated, the benchmark has not been executed, and the governance protocol has not been piloted in a live mathematical project.

\section{Prospective understanding audit}

The unit of audit may be one person or a named group whose distributed competencies are documented. Each dimension may be scored 0 (not demonstrated), 1 (partial or heavily scaffolded), or 2 (independently demonstrated). Scores are descriptive. No pass threshold is validated.

\begin{longtable}{p{0.06\linewidth}p{0.22\linewidth}p{0.61\linewidth}}
\toprule
ID & Dimension & Challenge \\
\midrule
UA1 & Strategy & State the problem, central strategy, and purpose of main definitions without reproducing the artifact. \\
UA2 & Dependencies & Identify assumptions, imported results, formal dependencies, and model-generated steps. \\
UA3 & Reconstruction & Reconstruct decisive transitions at field-appropriate granularity. \\
UA4 & Error diagnosis & Find or explain a planted error and trace its consequences. \\
UA5 & Transfer & Apply the strategy to a nearby case or explain why transfer fails. \\
UA6 & Significance & Explain novelty, limitations, and relation to prior work. \\
UA7 & Authorization & Decide whether to release, revise, or withdraw, and accept responsibility for the decision. \\
\bottomrule
\end{longtable}

Recommended administration is delayed, transcript-restricted, and partly adversarial. The audit record should identify which member satisfied each dimension, allowed tools, challenge prompts, assessor, and unresolved disagreements. A participant may use AI while learning; the test asks whether reconstructible competence and authority remain available when the artifact is challenged.

\section{Contribution-authority trace}

For contributor $i$, record $C_i=(P_i,E_i)$, where $P_i$ is productive contribution and $E_i$ is the scope of epistemic authority. Productive importance does not automatically confer authority, and authority does not erase infrastructural contribution.

\begin{longtable}{p{0.24\linewidth}p{0.19\linewidth}p{0.19\linewidth}p{0.29\linewidth}}
\toprule
Role & Typical $P$ & Typical $E$ & Responsibility boundary \\
\midrule
Conceptual lead & High & High & Problem, strategy, and significance. \\
Proof formalizer & High & Medium & Encoding and local correctness. \\
Library maintainer & High & Low & Infrastructure integrity and releases. \\
Model operator & Medium & Low & Accurate trace and disclosure. \\
Independent reviewer & Medium & High & Acceptance recommendation and challenge. \\
Expositor & Medium & Medium & Clarity and faithful reconstruction. \\
Generative model & Variable & None & No testimonial or moral responsibility. \\
Authorizing human & Variable & High & Release, revision, stop, and withdrawal. \\
\bottomrule
\end{longtable}

\section{Optional FCA analyses}

The disclosed binary context contains seven objects and eight attributes. Standard enumeration produced 13 formal concepts. The fuzzy view uses Jaccard similarity only as a sensitivity-oriented representation. Its strongest pairs were collaborative/governed studio (0.75), formal-computational/governed studio (0.75), and degenerate studio/fact-graph orchestration (0.75). These values reflect the hand-coded context.

The stochastic view used 2,000 replicates, seed 25082026, and independent 0.05 cell-flip probability. The strongest simple implications were understanding audit $\Rightarrow$ human judgment (0.797), understanding audit $\Rightarrow$ public accountability (0.7905), understanding audit $\Rightarrow$ human stop right (0.788), and formal verification $\Rightarrow$ repository memory (0.785). These numbers estimate robustness to a specified coding perturbation. They are not probabilities that the ontology or implications are true.

\section{Robustness and heterogeneity map}

The RHM uses three statuses:

\begin{itemize}
\item \texttt{ROBUST\_CORE}: survives the present source, boundary, and counterargument checks at its stated claim type.
\item \texttt{SPURIOUS\_ENVELOPE}: depends on role confusion, unsupported promotion, unstable coding, or a claim stronger than its evidence.
\item \texttt{LEGITIMATE\_OPEN\_UNCERTAINTY}: remains meaningful and testable but is not resolved by current evidence.
\end{itemize}

Heterogeneity should be tracked across mathematical field, task type, career stage, formalization maturity, model access, language, institutional resources, and time horizon. An aggregate result should not be promoted when opposite effects across these strata are plausible.

\section{Prospective validation and extension program}

The extension is staged so that scale does not precede validity. The machine-readable roadmap is \texttt{data/extension\_roadmap.csv}.

\begin{longtable}{p{0.10\linewidth}p{0.22\linewidth}p{0.42\linewidth}p{0.18\linewidth}}
\toprule
Stage & Objective & Design and primary outputs & Stop rule \\
\midrule
I & Validate the understanding and governance instruments. & A manually assessed pilot with independent coders, delayed reconstruction, planted-error diagnosis, transfer, and claim-relative GR1--GR5 coding. & Do not compare workflows if inter-rater reliability or construct validity is inadequate. \\
II & Compare workflows prospectively. & Authorial, assistive ($L1$--$L2$), governed studio ($L3$--$L4$), and degenerate studio ($L4$--$L5$) conditions across proof construction, formalization, counterexample search, and exposition. & Freeze tasks, criteria, and instruments before outcome inspection; retain null or adverse results. \\
III & Pilot corpus classification. & A stratified gold set of 100--200 arXiv papers, independently double-coded by field, year, contribution type, and declared tool use. & Do not scale if human-human and model-human agreement are below preregistered thresholds or show major domain drift. \\
IV & Scale the arXiv study. & A 500--1,000 paper pilot followed, if warranted, by a larger longitudinal corpus with uncertainty intervals and field-specific error analysis. & Do not interpret missing disclosures as absence of AI use; do not pool strata with incompatible error profiles. \\
\bottomrule
\end{longtable}

For Stage II, primary outcomes are delayed reconstruction, error diagnosis, transfer, correctness, originality, time, and calibration. Secondary outcomes include perceived agency, dependence on proprietary infrastructure, reviewer burden, and distributional effects by career stage. Tasks, assessors, and planted errors should be blinded where feasible. Model versions and material prompts should be archived subject to licensing and privacy. The calibration tasks must remain separate from the final evaluation set.

For Stages III and IV, the codebook must be frozen before large-scale classification. The gold set should include positive, negative, ambiguous, and missing-disclosure cases. Domain-specific confusion matrices, not one aggregate accuracy value, determine whether scaling is defensible. Historical examples must not be recoded as prospective observations, and corpus prevalence must not be interpreted as causal evidence about understanding or institutional effects. This supplement specifies the program but reports no validation or benchmark result.

\section{Figures and alt text}

\begin{figure}[p]
\centering
\includegraphics[width=\textwidth]{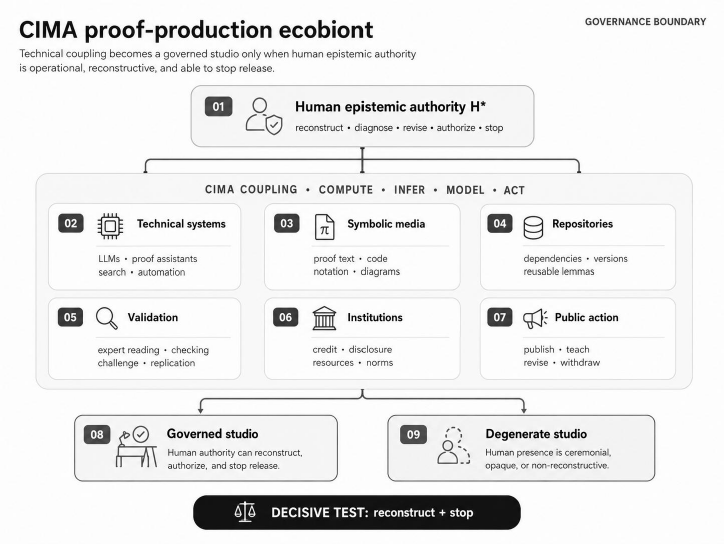}
\caption{CIMA proof-production ecobiont and the governance boundary.}
\end{figure}
\paragraph{Alt text.} A human authority node $H^{*}$ connects to technical systems, symbolic media, repositories, validation, institutions, and public action. The coupled system branches to a governed studio when humans can reconstruct and stop release, and to a degenerate studio when oversight is ceremonial.

\begin{figure}[p]
\centering
\includegraphics[width=\textwidth]{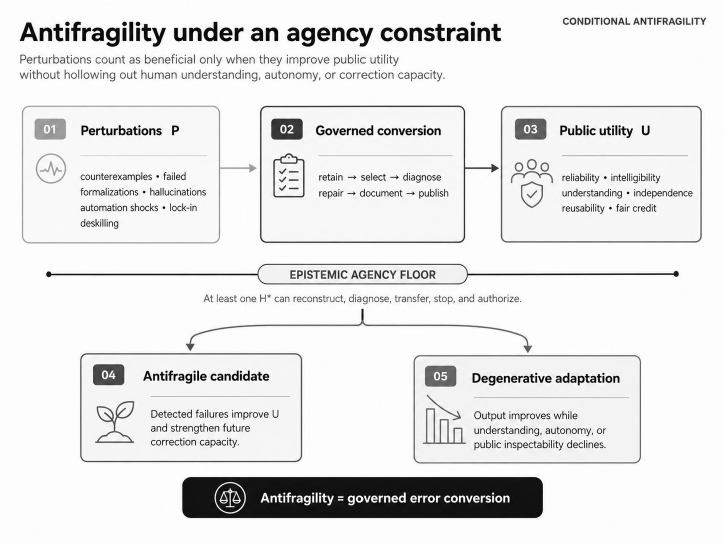}
\caption{Conditional antifragility with an epistemic agency constraint.}
\end{figure}
\paragraph{Alt text.} Perturbations pass through governed conversion into public utility. A horizontal agency constraint separates an antifragile candidate, where correction capacity improves, from degenerative adaptation, where output improves while understanding or autonomy declines.

\begin{figure}[p]
\centering
\includegraphics[width=\textwidth]{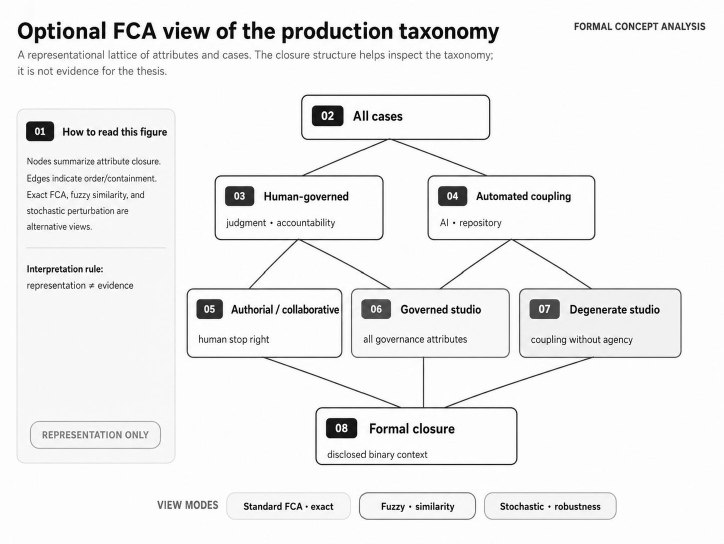}
\caption{Optional FCA view of the production taxonomy.}
\end{figure}
\paragraph{Alt text.} A simplified lattice-like arrangement distinguishes human-governed and automated coupling branches, with authorial/collaborative, governed-studio, and degenerate-studio cases. A note states that the view is a representation and not evidence.

\begin{figure}[p]
\centering
\includegraphics[width=\textwidth]{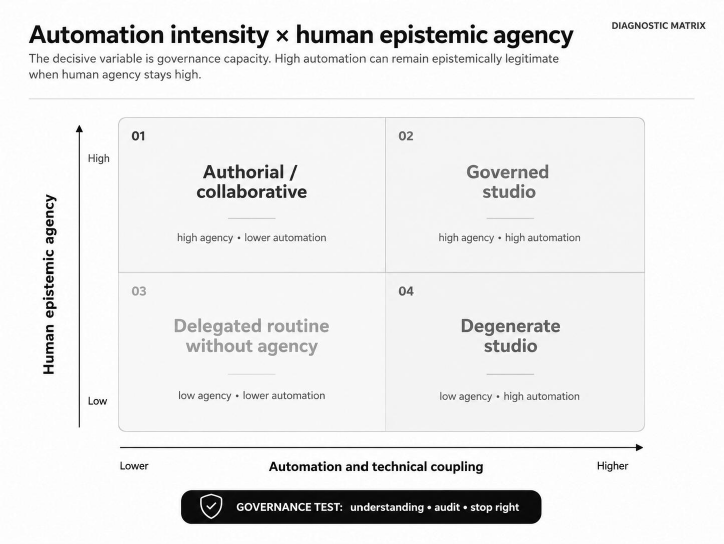}
\caption{Automation intensity and human epistemic agency.}
\end{figure}
\paragraph{Alt text.} A two-by-two matrix places authorial or collaborative practice at high agency and lower automation, governed studios at high agency and high automation, delegated routine at low agency and lower automation, and degenerate studios at low agency and high automation.

\begin{figure}[p]
\centering
\includegraphics[width=\textwidth]{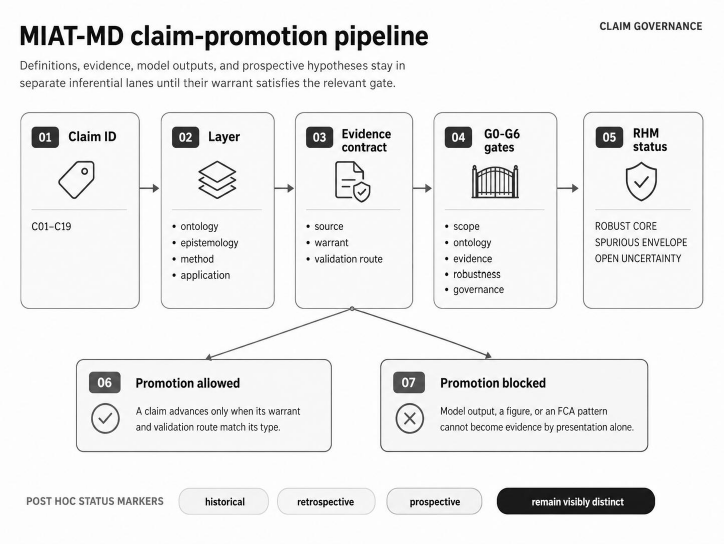}
\caption{MIAT-MD claim-promotion pipeline.}
\end{figure}
\paragraph{Alt text.} Claims move from stable IDs through layer classification, evidence contracts, G0--G6 gates, and RHM status. A green branch allows promotion when warrant matches claim type; a red branch blocks promotion from model output, figures, or FCA patterns alone.

\begin{figure}[p]
\centering
\includegraphics[width=\textwidth]{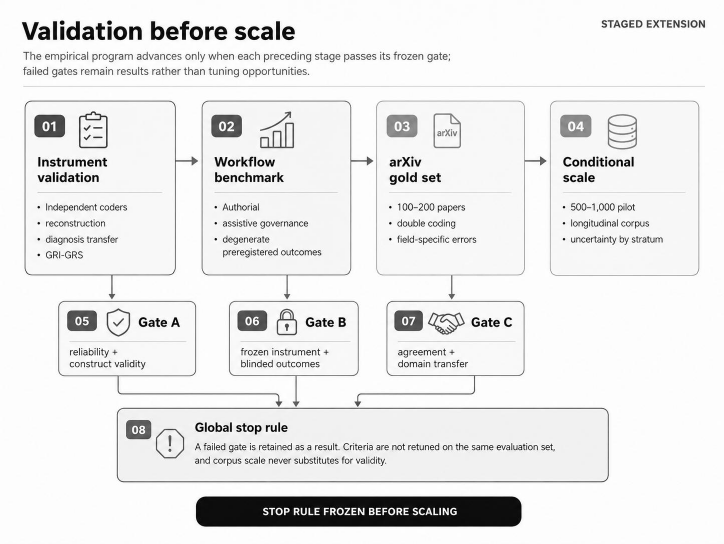}
\caption{Staged validation and extension sequence.}
\end{figure}
\paragraph{Alt text.} Four stages move from instrument validation to workflow comparison, a manually double-coded arXiv gold set, and conditional corpus scaling. Gates between stages stop progression when reliability, construct validity, or domain-transfer criteria fail.

\section{Tool and computational resource disclosure for this release}

Generative AI systems supported literature triage, argument mapping, language revision, LaTeX editing, and code generation. The human author defined the research question and framework, selected the claims advanced, and retains responsibility for the interpretations, references, classifications, and final text. Because the manuscript is conceptual rather than a reported mathematical proof, the $L0$--$L5$ proof-intervention scale does not map perfectly onto its workflow. The closest description is editorial and analytic assistance at approximately $L1$--$L2$, with material use disclosed here. No benchmark, participant study, or corpus analysis was generated or represented as completed.

\section{Reproducibility manifest}

The arXiv submission bundle for version 2.6.1 is intentionally minimal: this single \LaTeX{} source and the six PDF figures it loads. A separate, versioned reproducibility archive is available at \url{https://drive.google.com/file/d/1IlBFXJd1z1-F6xjPO0ITQT3j-o_FvgBP/view}. That archive contains the compiled manuscript, its single-file source with this supplement embedded, claim, case, decision, governance, and extension CSV ledgers, analysis scripts, the figure-generation script, JSON/CSV outputs, figure files in PNG/SVG/PDF, a spreadsheet mirror, an environment freeze, and SHA-256 checksums. Standard FCA, fuzzy similarity, stochastic perturbation, and RHM outputs can be regenerated without network access. The PDF build requires a standard \LaTeX{} installation.

\end{document}